\documentclass{aastex631}
\newcommand{\alphadead}{\alpha_{\rm dead}}
\newcommand{\alphamri}{\alpha_{\rm mri}}
\newcommand{\amin}{a_{\rm min}}
\newcommand{\amax}{a_{\rm max}}
\newcommand{\aexp}{a_{\rm exp}}

\newcommand{\firr}{F_{\rm irr}}
\newcommand{\hgas}{H_{\rm gas}}
\newcommand{\mstar}{M_\ast}
\newcommand{\mdisko}{M_{\rm disk,0}}
\newcommand{\mdot}{\dot{M}}
\newcommand{\mdotouter}{\dot{M}_{\rm 3AU}}
\newcommand{\p}{\partial}
\newcommand{\rstar}{R_\ast}
\newcommand{\stefan}{\sigma_{\rm SB}}
\newcommand{\st}{{\rm St}}
\newcommand{\tmri}{T_{\rm mri}}
\newcommand{\taudiff}{\tau_{\rm diff}}
\newcommand{\tauheat}{\tau_{\rm heat}}
\newcommand{\tauspread}{\tau_{\rm spread}}
\newcommand{\tstar}{T_\ast}
\newcommand{\uwind}{u_{\rm wind}}

\shorttitle{Large Fluctuations within 1 AU in Protoplanetary Disks}
\shortauthors{Chambers}

\begin{document}

\title{Large Fluctuations within 1 AU in Protoplanetary Disks}

\author{John Chambers}
\affiliation{Carnegie Institution for Science \\
5241 Broad Branch Road NW \\
Washington, DC 20015, USA}

\begin{abstract}
Protoplanetary disks are often assumed to change slowly and smoothly during planet formation. Here, we investigate the time evolution of isolated disks subject to viscosity and a disk wind. The viscosity is assumed to increase rapidly at around 900 K due to thermal ionization of alkali metals, or thermionic and ion emission from dust, and the onset of magneto-rotational instability (MRI). The disks generally undergo large, rapid fluctuations for a wide range of time-averaged mass accretion rates. Fluctuations involve coupled waves in temperature and surface density that move radially in either direction through the inner 1.5 AU of the disk. Two types of wave are seen with radial speeds of roughly 50 and 1000 cm/s respectively. The pattern of waves repeats with a period of roughly 10,000 years that depends weakly on the average mass accretion rate. Viscous transport due to MRI is confined to the inner disk. This region is resupplied by mass flux from the outer disk driven by the disk wind. Interior to 1 AU, the temperature and surface density can vary by a factor of 2--10 on timescales of years to ky. The stellar mass accretion rate varies by 3 orders of magnitude on a similar timescale. This behavior lasts for at least 1 My for initial disks comparable to the minimum-mass solar nebula.

\end{abstract}

%
%
\section{Introduction}
Protoplanetary disks provide the setting for planet formation, but they are not simply a passive backdrop. Disks affect many of the processes involved in planetary growth. Dust coagulation, planetesimal formation, pebble accretion, oligarchic growth, and planetary migration are all influenced by the conditions in the disk such as the temperature and gas pressure \citep{ida:1993, paardekooper:2009, birnstiel:2010, ormel:2010, simon:2016}.

Most of these processes take place on timescales that are short compared to the lifetime of a typical disk. As a result, it is common to assume that the disk properties remain constant or change slowly and smoothly over time. If disks undergo rapid changes, instead, these variations will have to be taken into account when considering the growth of planets.

The dominant mechanism that drives the evolution of protoplanetary disks is uncertain. Until recently, turbulence generated by the magneto-rotational instability (MRI) was the leading candidate for transporting mass and angular momentum through disks \citep{balbus:1991}. The viscosity associated with this turbulence would release orbital energy, providing an important source of heating, especially in the inner disk \citep{pringle:1981}.

MRI requires a certain degree of ionization in the gas in order to operate. Close to the star, collisional ionization of alkali metals may be sufficient to activate MRI in a protoplanetary disk \citep{gammie:1996}. Alternatively, MRI initiation may be set by thermionic and ion emission from dust grains, although the temperature dependence is similar \citep{desch:2015}. However, much of the rest of the disk is probably a ``dead zone'' where MRI is suppressed \citep{bai:2013}. Other, non-magnetic, sources of turbulence may exist, but these are likely to cause much weaker transport and heating than MRI \citep{latter:2022}.

Disk winds provide a plausible alternative evolutionary mechanism \citep{bai:2016, suzuki:2016}. A wind removes mass from the disk, and the escaping mass takes more than its fair share of angular momentum with it. Material that remains in the disk flows inwards, driving disk evolution. Unlike MRI, disk winds are expected to generate little heating or turbulence, especially near the disk midplane \citep{mori:2019}. In regions where winds always dominate, it is plausible that the disk evolves gradually.

However, because MRI produces heat and requires a minimum temperature to operate, the evolution is susceptible to a positive feedback. It is likely that some parts of a disk can exist in two temperature states---one dead and one active---for conditions that are otherwise identical. In these circumstances, instabilities can occur, and the assumption that the disk evolves slowly and smoothly over time may not be correct.

In this paper, we examine the time evolution of isolated protoplanetary disks with masses similar to the minimum mass of the solar nebula. We show that the inner regions (roughly $<1$ AU) are typically unstable for a wide range of disk mass accretion rates. These disks undergo large, rapid changes that repeat roughly every $10^4$ years. In Section~2, we describe the disk model used. In Section~3, we briefly discuss relevant evolutionary timescales. Section~4 examines steady-state evolution and its stability. Section~5 considers time evolution calculated numerically. In Section~6, we discuss some implications, while Section~7 contains a summary.

%
%
\section{Disk Model}
We consider a 1D radial model for a thin, axisymmetric disk. Following \citet{hameury:1998, wu:2021}, the evolution of the midplane temperature $T$ can be calculated from energy balance as follows:
\begin{equation}
c_p\Sigma\frac{\p T}{\p t}=
\frac{9\Sigma\nu\Omega^2}{4}
+\frac{2\firr}{(1+3\tau/4)}
-\frac{2\stefan T^4}{(1+3\tau/4)}
+\frac{3}{a}\frac{\p}{\p a}
\left[a\nu c_p\Sigma\frac{\p T}{\p a}\right]
\label{eq-dtdt}
\end{equation}
where $a$ is radial distance from the star, $\Omega$ is the orbital angular velocity, $\Sigma$ is the surface density, $\nu$ is the viscosity, $c_p$ is the specific heat capacity, $\stefan$ is the Stefan-Boltzmann constant, and $\tau$ is the vertical optical depth of the midplane, given by 
\begin{equation}
\tau=\frac{\kappa\Sigma}{2}
\end{equation}
where $\kappa=5$ cm$^2$/g is the opacity per unit mass of gas, which is an approximate, temperature-averaged value of more realistic opacity estimates \citep{pollack:1994}.

The first and second terms on the righthand side of Eqn.~\ref{eq-dtdt} are the heating rates per unit area due to viscous accretion and stellar irradiation. The third term is the cooling rate from the disk surfaces. The last term represents radial diffusion of heat due to turbulent eddies. Note that this is typically more important than radial radiative thermal diffusion \citep{ludwig:1998}.

Following \cite{chiang:1997}, the stellar irradiation flux is approximately given by
\begin{equation}
\firr=0.1\left(\frac{\rstar}{a}\right)^3\stefan\tstar^4
+\left(\frac{2}{7}\right)^{8/7}
\left(\frac{k\rstar\tstar}{G\mstar\mu m_H}\right)^{4/7}
\left(\frac{\rstar}{a}\right)^{12/7}
\stefan\tstar^4
\end{equation}
where $k$ is Boltzmann's constant, $m_H$ is the mass of a hydrogen atom, $\mu$ is the mean molecular weight of the gas, and $\mstar=M_\odot$ is the stellar mass. Also, $\rstar=0.01$ AU is the stellar radius, and $\tstar=4200$ K is the stellar temperature, which are typical values for a solar-mass pre-main sequence star \citep{choi:2016}. The first and second terms on the righthand side of this equation represent the limiting cases where the physical size of the star is important, and where it can be treated as a point source, respectively.

Following \citet{suzuki:2016}, we consider the surface density evolution due to viscosity and a disk wind, such that
\begin{equation}
\frac{\p\Sigma}{\p t}=
\frac{3}{a}\frac{\p}{\p a}\left[
a^{1/2}\frac{\p}{\p a}(a^{1/2}\nu\Sigma)\right]
-\frac{1}{a}\frac{\p}{\p a}(a\uwind\Sigma)
\label{eq-dsdt}
\end{equation}
where $\uwind$ is the radial velocity of the gas induced by the loss of angular momentum due to the disk wind. We include a disk wind to ensure that the disk continues to evolve in regions where viscosity is low. The precise behavior of the wind is not important here, and we neglect a loss term due to entrainment of disk mass in the wind.

We use an $\alpha$ model for the viscosity, where
\begin{equation}
\nu=\frac{\alpha c_s^2}{\Omega}
\end{equation}
where $c_s$ is the sound speed. Following \citet{latter:2012}, we assume that $\alpha$ can be expressed as a function of $T$ such that
\begin{eqnarray}
\alpha=\left(\frac{\alphamri+\alphadead}{2}\right)
+\left(\frac{\alphamri-\alphadead}{2}\right)
\tanh\left(\frac{T-\tmri}{\rm \Delta\tmri}\right)
\label{eq-alphat}
\end{eqnarray}
where $\tmri=900$ K is the temperature at which MRI becomes effective, and $\Delta\tmri=150$ K, with these values adopted from \citet{latter:2012}. In addition, $\alphamri=10^{-2}$ and $\alphadead=10^{-4}$. Note that we allow for a small amount of turbulence in the dead zone due to sources other than MRI.

The equations for $\Sigma$ and $T$ are solved using an explicit scheme with an adaptive timestep that is chosen to keep the maximum change per step below 1\%. Typical values of the timestep are less than or comparable to 1 day when part of the disk is in the high-$\alpha$ state. The timestep is larger when the entire disk is in the low-$\alpha$ state.

%
%
\section{Timescales}
We can get some insights into the dynamical evolution of the disk by examining a few key timescales. In particular, we consider the viscous heating timescale $\tauheat$, viscous spreading timescale $\tauspread$, and radial thermal diffusion timescale $\taudiff$. These are given by
\begin{eqnarray}
\tauheat&=&T\left(\frac{\p T}{\p t}\right)_{\rm visc}^{-1}
\sim\frac{1}{\alpha\Omega}
\nonumber \\
\tauspread&=&\Sigma\left(\frac{\p\Sigma}{\p t}\right)_{\rm visc}^{-1}
\sim\frac{1}{\alpha\Omega}
\left(\frac{a}{\hgas}\right)^2
\left[\frac{\p\ln a}{\p\ln(\nu\Sigma)}\right]^{2}
\nonumber \\
\taudiff&=&T\left(\frac{\p T}{\p t}\right)_{\rm diff}^{-1}
\sim\frac{1}{\alpha\Omega}
\left(\frac{a}{\hgas}\right)^2
\left[\frac{\p\ln a}{\p\ln(\nu\Sigma)}\right]
\left[\frac{\p\ln a}{\p\ln T}\right]
\end{eqnarray}
where the suffixes visc and diff refer respectively to the viscous and radial diffusion terms in Eqns.~\ref{eq-dtdt} and \ref{eq-dsdt}. We have assumed that $c_p=7k/2\mu m_H$.

In a disk with shallow viscosity and surface density profiles, $\tauheat$ is short compared to $\tauspread$. This justifies the common procedure in which the heating and cooling are assumed to balance, and the equilibrium temperature is used to calculate the surface density evolution. For shallow profiles, $\tauheat$ is also short compared to $\taudiff$, which explains why it is often assumed that the temperature at each radial location can be calculated independently of other locations.

We note, however, that both of these approximations break down if the viscosity and surface density profiles become steep enough. In this case, $T$ and $\Sigma$ can change at similar rates, and radial heat flows may be important.

\begin{figure}
\begin{center}
\includegraphics[height=12cm]{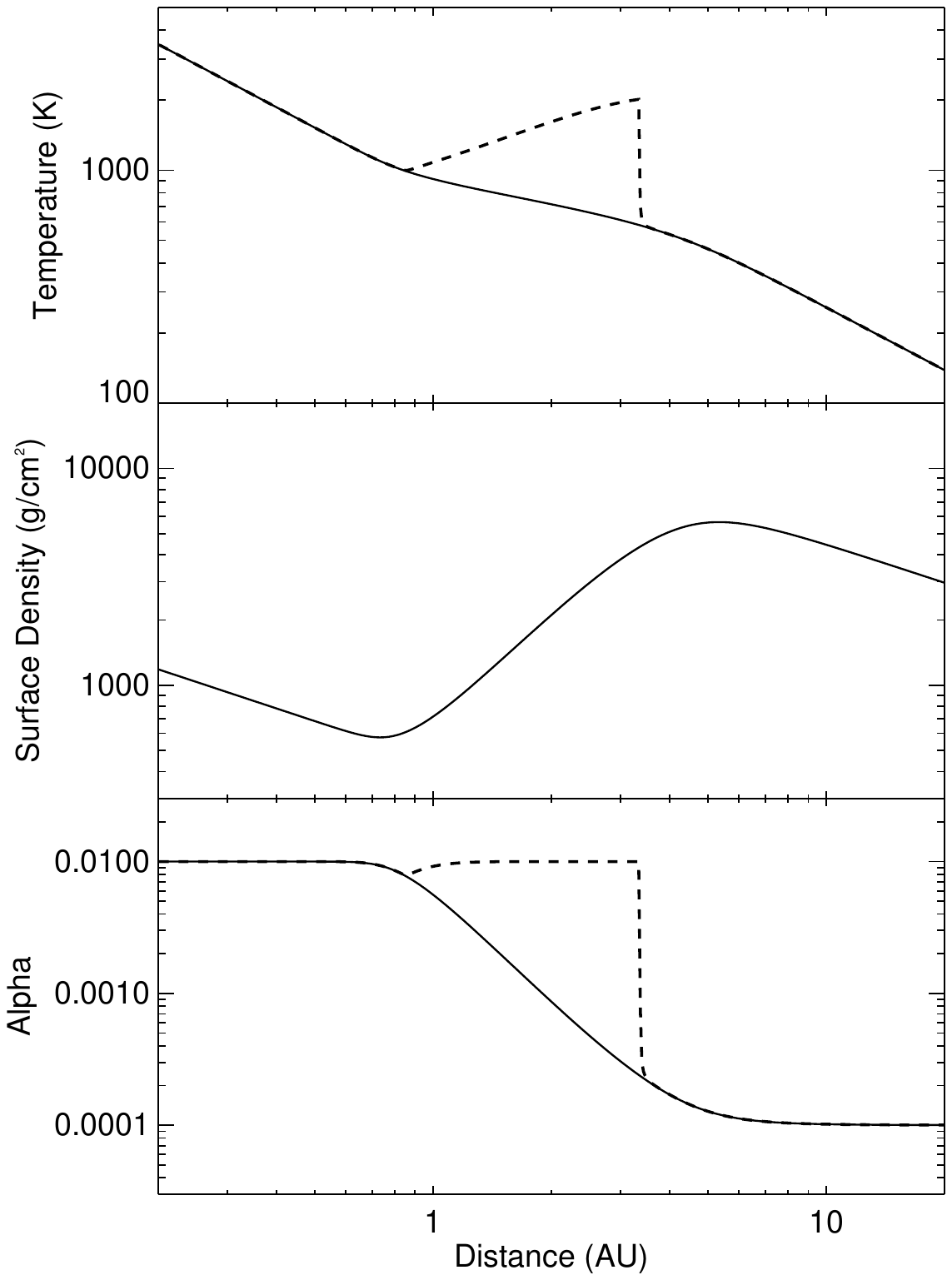}
\end{center}
\caption{Solid curves: steady-state profiles for temperature (top panel), surface density (middle) and $\alpha$ (bottom) in a disk with a mass flux of $10^{-7}M_\odot$/y. Dashed curves: the disk after the temperature is allowed to evolve for 30,000 years with the surface density held fixed.}
\end{figure}

%
%
\section{Steady State Approximation}
A common approach to modeling a disk is to assume that the inward mass flux through the disk is independent of distance from the star \citep{terquem:2008, kretke:2010, mohanty:2018, jankovic:2021}. This ``steady state'' approximation can be reasonable in the inner regions of smooth, viscous disks where the viscous timescale is short compared to the age of the disk.

As an example, we consider a simplified model in which viscous heating and cooling are balanced, and viscosity is the only factor affecting the surface density evolution. The midplane temperature in this case is given by
\begin{equation}
T=\left(\frac{27k\kappa\Sigma^2\alpha\Omega}
{64\stefan\mu m_H}\right)^{1/3}
\label{eq-tmid-steady}
\end{equation}
where we have assumed that $\tau\gg1$, and the opacity $\kappa$ is assumed to be a constant.

Steady state implies that the mass flux $\mdot=3\pi\nu\Sigma$, where $\mdot$ is a constant. Thus, we get
\begin{equation}
T^5=\frac{3\mu m_H\kappa\mdot^2\Omega^3}
{64\pi^2k\stefan\alpha}
\end{equation}
with the corresponding surface density proflie obtained from Eqn.~\ref{eq-tmid-steady}.

The solid curves in Fig~1 shows the steady state profiles for $T$, $\Sigma$ and $\alpha$ in a case with $\mdot=10^{-7} M_\odot$/y. Far from the transition point $\tmri$, the temperature falls steeply with distance. The surface density profile has to be quite shallow in these regions in order to maintain a fixed mass flux. In the region close to $\tmri$, $\alpha$ undergoes a rapid transition from $\alphamri$ to $\alphadead$. The decrease in $\alpha$ buffers the temperature, leading to a shallow $T$ profile. Conservation of $\mdot$ means that $\Sigma$ increases with distance in this region.

\begin{figure}
\begin{center}
\includegraphics[height=12cm]{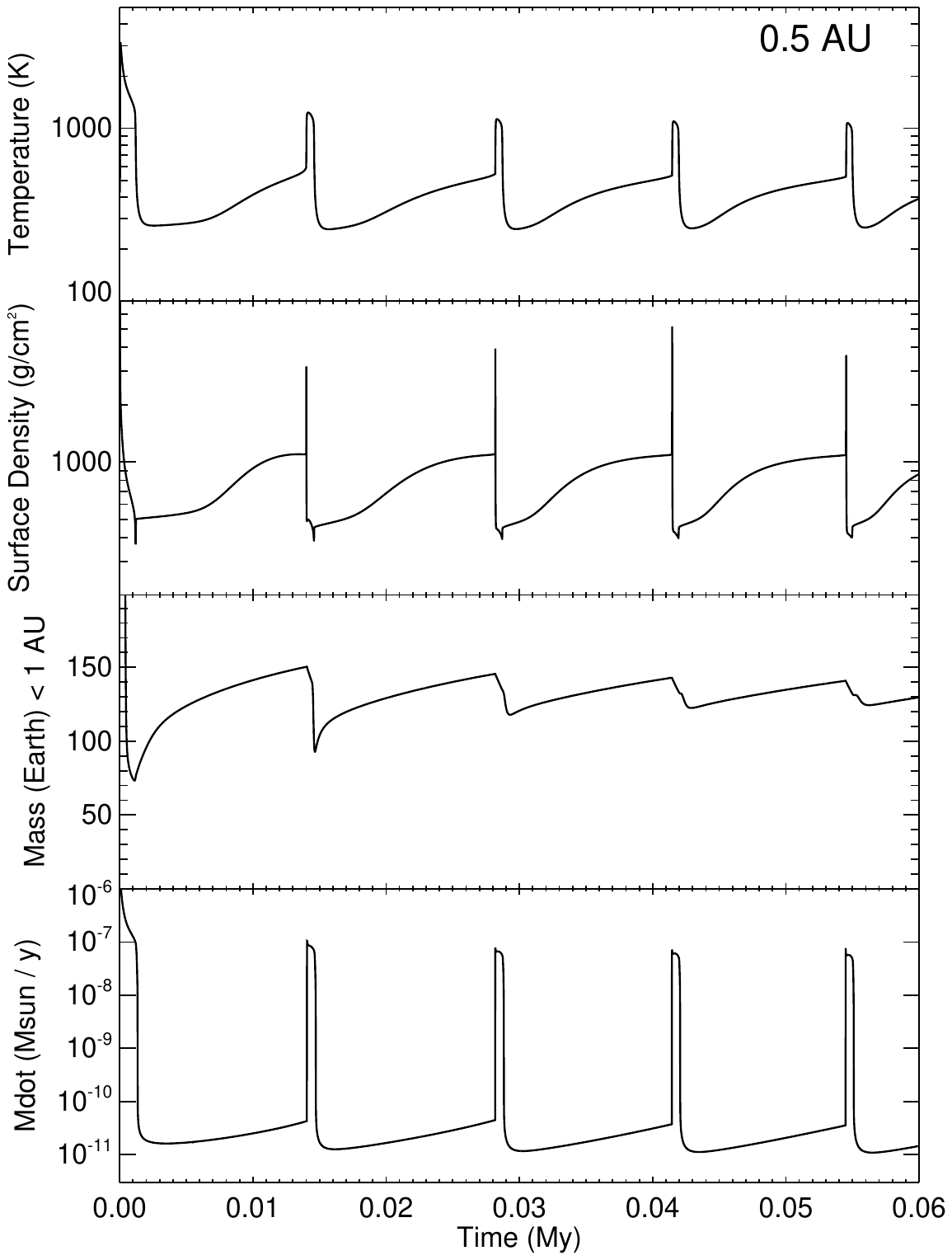}
\end{center}
\caption{Evolution of a disk undergoing viscous accretion and viscous heating only. The top two panels show the temperature and surface density at 0.5 AU from the star. The third panel shows the total mass interior to 1 AU. The bottom panel shows the accretion rate onto the star.}
\end{figure}

%
%
\subsection{Is the Steady State Case Stable?}
In this section, we examine the stability of the steady-state solution obtained above. We will show that this solution is typically unstable when $\alpha$ varies rapidly with temperature despite the fact that steady-state solutions are widely used.

We begin with the equilibrium temperature given by
\begin{equation}
T_0=\left(\frac{27k\kappa\Sigma_0^2\alpha_0\Omega}
{64\stefan\mu m_H}\right)^{1/3}
\end{equation}
where the subscript 0 indicates the equilibrium value of a quantity.

Consider a small temperature perturbation $T=T_0+\delta T$. Since temperature typically changes faster than surface density, we hold $\Sigma$ fixed. Using the appropriate terms in Eqn.~\ref{eq-dtdt}, the perturbation causes the temperature to change at a rate given by
\begin{equation}
\frac{dT}{dt}=
\left(\frac{9k\Omega}{4\mu m_Hc_p}\right)
(T_0+\delta T)\left(\alpha_0+\frac{d\alpha}{dT} \delta T\right)
-\left(\frac{16\stefan}{3\kappa c_p\Sigma_0^2}\right)
(T_0^4+4T_0^3\delta T)
\end{equation}

After subtracting the steady-state solution, this leads to
\begin{equation}
\left(\frac{4\mu m_H}{9k\Omega}\right)
\frac{c_p}{\alpha_0}\frac{dT}{dt}=
\left(\frac{d\ln\alpha}{d\ln T}-3\right) \delta T
\end{equation}

Thus, if $\alpha$ is a steeper function of temperature than $T^3$, a small perturbation will grow larger. In the transition region between the dead zone and active MRI, $\alpha$ is likely to depend much more steeply on $T$ than this. (In Eqn.~\ref{eq-alphat}, for example, the exponent is as large as 8.) Thus, $T$ and $\alpha$ will evolve towards values set by either $\alphadead$ or $\alphamri$.

To see this in practice, we calculate the temperature evolution numerically, beginning with the steady-state solution shown by the solid curve in Fig~1, and holding $\Sigma$ fixed. The numerical scheme introduces small errors that are amplified by the instability described above. After $3\times10^4$ years, the system has evolved to the state shown by the dashed curve in Fig.~1. At this point, $\alpha$ is close to one of the two limiting values everywhere in the disk. The temperature profile has also changed substantially, with an abrupt transition near 3 AU.

\citet{mohanty:2018} carried out a similar stability analysis for steady-state solutions for MRI-powered disks in which the viscosity increases rapidly with temperature. They found that their steady-state solutions were typically unstable, and suggested that the disk would break up into rings as a result. This is partially correct, but the behavior turns out to be more complicated as we will see below.

As we saw in Section 3, a steep gradient in $T$ is likely to cause rapid changes in $\Sigma$ on a similar timescale to changes in $T$. Thus, the dashed-line solution in Fig.~1 will evolve once $\Sigma$ is allowed to vary. Radial diffusion of heat may also be important. In order to determine how the disk actually behaves, it is necessary to calculate the time evolution of $T$ and $\Sigma$ simultaneously, and we do this in the next section.

\begin{figure}
\begin{center}
\includegraphics[height=12cm]{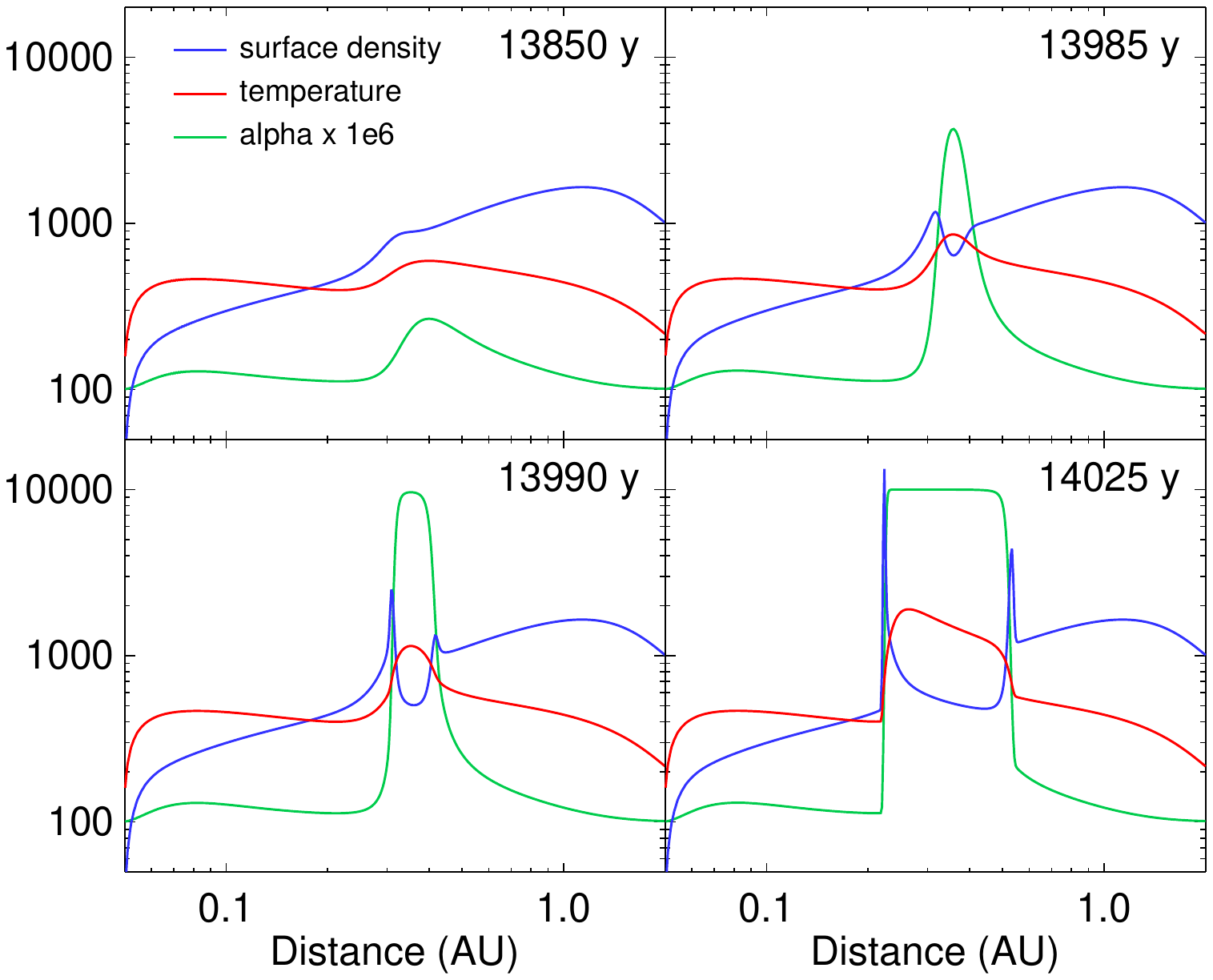}
\end{center}
\caption{Four snapshots of the inner portion of the disk seen in Fig~2 showing the onset of waves that travel through the disk. The red curves show temperature in K. The blue curves show surface density in g/cm$^2$.  The green curves show $\alpha$ multiplied by $10^6$.}
\end{figure}

%
%
\section{Time Varying Case}
In this section, we calculate the coupled evolution of Eqns.~\ref{eq-dtdt} and \ref{eq-dsdt} numerically. To begin with, we neglect the disk wind term in Eqn.~\ref{eq-dsdt}, and the stellar irradiation term in Eqn.~\ref{eq-dtdt}, but we include radial thermal diffusion. We use a grid with 1000 logarithmically spaced cells between $\amin=0.05$ AU and $\amax=200$ AU. The initial temperature and surface density profiles are
\begin{eqnarray}
T&=&T_i\left(\frac{a}{\rm 1\ AU}\right)^{-1/2} 
\nonumber \\
\Sigma&=&\Sigma_i\left(\frac{a}{\rm 1\ AU}\right)^{-1} 
\exp\left(-\frac{a}{\aexp}\right)
\label{eq-profiles}
\end{eqnarray}
where $T_i=300$ K and $\aexp=30$ AU, and $\Sigma_i$ is chosen so that the total initial disk mass $\mdisko=0.04 M_\odot$.

Figure 2 shows how the temperature and surface density at 0.5 AU change over time. Also shown are the total mass interior to 1 AU, and the mass accretion rate onto the star. Instead of reaching a steady state, the evolution undergoes a repeated cycle with a period of roughly 14,000 years. At the start of each cycle, $T$ and $\Sigma$ gradually increase over time. They then undergo a brief spike and rapid decline before the cycle begins again. 

Over the course of one cycle, the temperature at 0.5 AU varies by a factor of about 4, while $\Sigma$ changes by an order of magnitude. The stellar mass accretion rate is even more dramatic, varying by almost 4 orders of magnitude. 

During each ``spike'' in Fig.~2, the temperature and surface density actually change smoothly. At 0.5 AU, $\Sigma$ takes about 5 years to rise to its maximum value, before falling again over the next 10 years. At the same location, $T$ takes about 50 years to go from minimum to maximum. Thus, changes that appear abrupt in Fig.~2 occur on a timescale that is an order of magnitude longer than the local orbital period.

Note that the simulation begins with smooth $T$ and $\Sigma$ profiles, following Eqns.~\ref{eq-profiles}. The region interior to 0.11 AU is initially hotter than 900~K, and thus MRI-active. No initial perturbation is required to initiate the cycles seen in Figure~2. In fact, large oscillations can occur in a disk that begins entirely in the MRI-dead state provided that mass inflow from the outer disk raises the temperature to $\sim 600$ K at some location. At this point, the positive feedback mechanism described in Section~4.1 leads to a rapid rise in $T$ and the onset of instability cycles.

\begin{figure}
\begin{center}
\includegraphics[height=12cm]{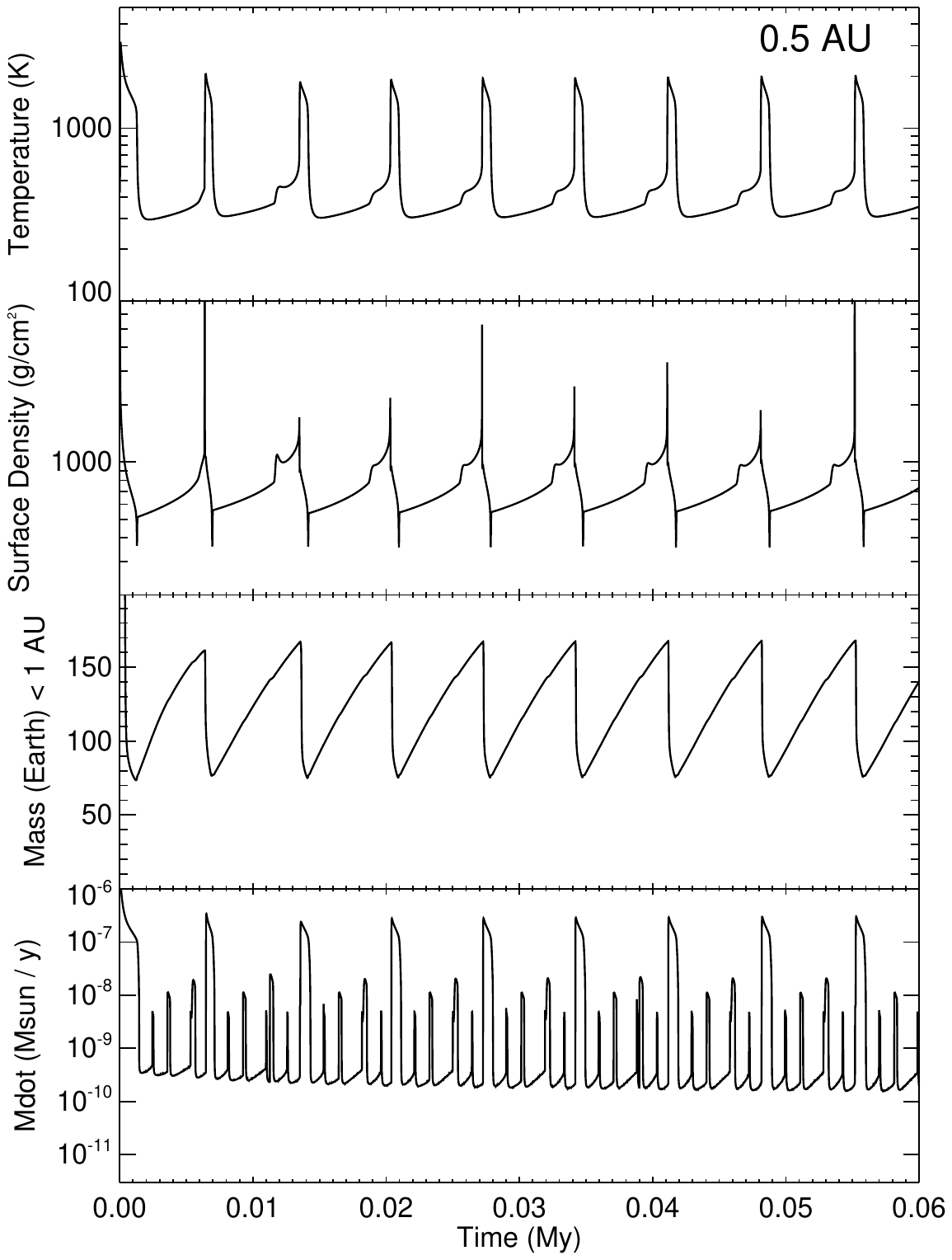}
\end{center}
\caption{Evolution of a disk when stellar irradiation and wind-driven transport are included. Other model details are identical to Fig.~2. The top two panels show the temperature and surface density at 0.5 AU from the star. The third panel shows the total mass interior to 1 AU. The bottom panel shows the accretion rate onto the star.}
\end{figure}

The epochs of rapid change are marked by coupled waves in $T$ and $\Sigma$ that sweep across the inner disk. Figure 3 shows the onset of two such waves. In the first panel in the figure, the entire disk is below 600 K, apart from a narrow region near 0.4 AU that is slightly warmer. (This structure is a remnant from the previous instability cycle.) This region undergoes a rapid, runaway increase in $T$, as seen in the second and third panels. The steep radial gradient in temperature at the inner edge of the hot region leads to a steep gradient in $\alpha$. The asymmetry in viscous transport across this gradient causes mass to pile up just interior to the temperature front. This generates a spike in $\Sigma$. Within the surface-density spike, the cooling rate is reduced due to increased optical depth, leading to a rapid rise in $T$. As a result, the temperature front moves inwards as a wave, as seen in the last panel of Fig.~3.

A second wave, moving in the opposite direction, can be seen at the outer edge of the hot region. This outer wave eventually stalls. However, the inner wave moves all the way to the inner edge of the disk. The temperature remains elevated for some time following the passage of the wave. This leads to an interval lasting $\sim 1000$ years marked by a high mass flux  onto the star. Eventually, the inner disk becomes depleted in mass and cools so that MRI ceases. Inflowing material gradually increases $\Sigma$ until the temperature becomes high enough for MRI to operate again, allowing new waves to form.

Figure 4 shows the evolution for an identical disk when we include stellar irradiation and transport driven by the disk wind with $\uwind=10$ cm/s. Here, the full versions of Eqns.~\ref{eq-dtdt} and \ref{eq-dsdt} are solved numerically. Both of these additions increase the mass flux flowing into the inner disk, with the disk-wind term making the dominant contribution.

Several differences from Fig.~2 are apparent. The temperature spikes in Fig.~4 are more pronounced, and the cycle period is reduced by about a factor of 2. The mass interior to 1 AU fluctuates by a larger amount in Fig.~4. The mass flux onto the star exhibits several secondary peaks in addition to the main spike at the end of each cycle. In addition, the minimum stellar accretion rate is an order of magnitude higher than in Fig.~2. These last two features are the result of additional, inward-moving waves (see below) that deposit a limited amount of mass onto the star in each case.

\begin{figure}
\begin{center}
\includegraphics[height=12cm]{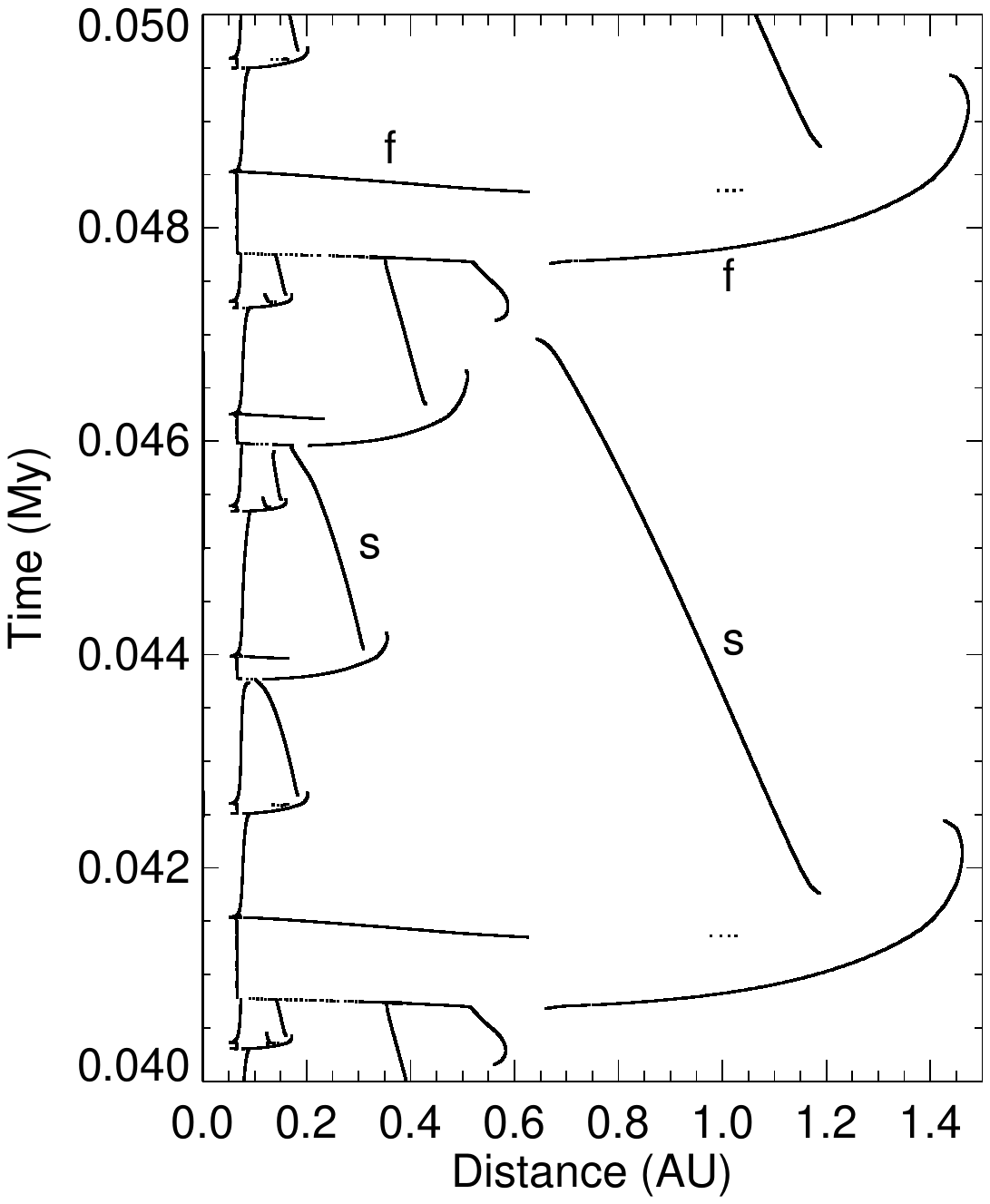}
\end{center}
\caption{Location of pressure maxima in the inner disk in the simulation shown in Fig.~4. Fast-moving waves associated with large $\alpha$, similar to those in Fig.~3, are denoted 'f'. Slower waves associated with small $\alpha$ are denoted `s'.}
\end{figure}

The simulation shown in Fig.~4 generates many temporary pressure maxima. These locations are of particular interest because they represent possible sites for dust accumulation and planetesimal formation. Figure 5 shows the radial locations of these pressure maxima during a 10,000-year time span, with maxima sampled at 1-year intervals. Pressure maxima appear and disappear, and move radially over time. The behavior is complicated, especially inside 0.5 AU, but two types of feature can be identified.

Fast-moving waves of the kind shown in Fig.~3 are common. Examples are indicated by the letter `f' in Fig.~5. These fast waves typically spawn inside 0.7 AU from the star. They can travel in either direction, at speeds $\sim 1000$ cm/s. Inward moving waves generally reach the inner edge of the disk, while outward moving waves stall at some point. No waves travel beyond 1.5 AU for the case shown here.

A second type of pressure-maximum wave is apparent, with examples indicated by the letter `s'. These features move more slowly than the fast waves, with speeds $\sim 50$ cm/s, and they generally move inwards. These structures represent the remnants of fast waves that have stalled, allowing the gas within them to cool so that $\alpha\sim\alphadead$. Slow waves also involve coupled oscillations in $T$ and $\Sigma$, but their features are less pronounced than fast waves and they more slowly since $\alpha$ is always small.

\begin{figure}
\begin{center}
\includegraphics[height=12cm]{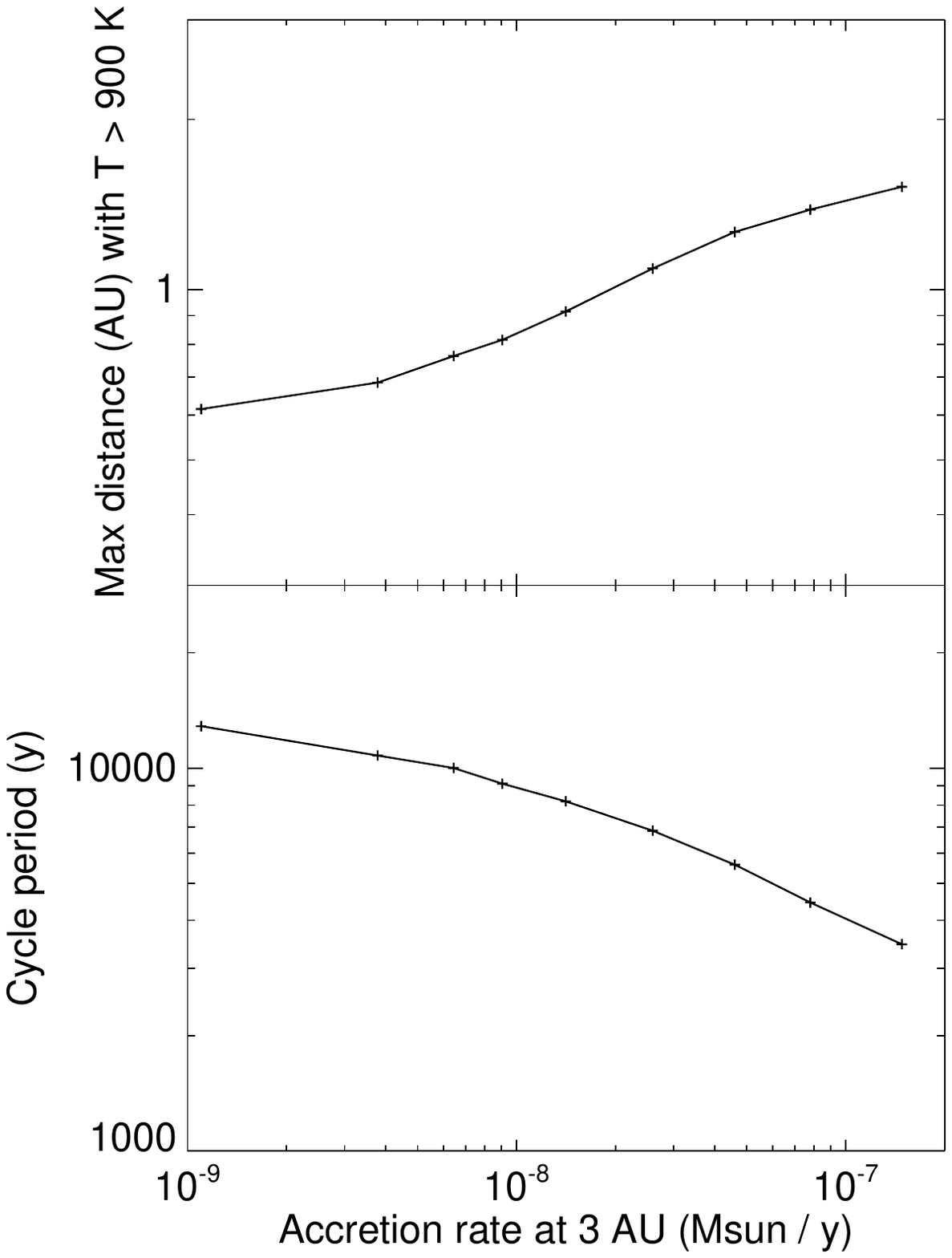}
\end{center}
\caption{The maximum distance at which the temperature exceeds 900 K (upper panel); and the period of temperature and surface density oscillations (lower panel) as a function of the inward mass flux at 3 AU.}
\end{figure}

The cyclic behavior seen in Figs.~2 and 4 suggests the evolution in the inner disk is controlled by conditions at larger distances. In particular, the mass flux entering the inner region is likely to play a central role. We can vary this flux by choosing a value of $\uwind$ that differs from the default value. To quantify the mass flux, we calculate the value $\mdotouter$ at 3 AU, which always lies beyond the unstable region. Both viscous and wind-driven flux contributions are included.

Figure 6 shows how two quantities vary with $\mdotouter$. The upper panel shows the maximum distance at which the temperature exceeds $\tmri=900$ K. This distance provides a rough measure of the radial extent of the instability cycles. Larger accretion rates lead to stronger viscous heating, so this distance increases with increasing $\mdotouter$, as one would expect. On long timescales, the mass flux will decline over time, and the portion of the disk that is unstable will shrink. After 3 My, for the simulation shown in Fig.~4, the instability cycles are confined to the inner 0.4 AU of the disk, for example. Even at this late stage, the steady-state solution is still not valid in this region.

The lower panel in Fig.~6 shows the period of the instability cycles in the inner disk. The cycles become faster as the accretion rate increases. We note that both quantities shown in Fig.~6 depend quite weakly on $\mdotouter$ for the disk parameters examined here.

%
%
\section{Discussion}
In the previous sections, we found that isolated protoplanetary disks comparable to the minimum-mass solar nebula undergo large fluctuations on timescales that are short compared to  a typical disk lifetime. Fluctuations consist of coupled temperature and surface density waves that travel radially through the inner 1.5 AU of disks for a wide range of disk mass accretion rates. Waves travel at speeds $\sim 50$--1000 cm/s, and can move in either direction. The pattern of waves approximately repeats on a timescale $\sim 10^4$ years.

This behavior is caused by a steep increase in viscosity with temperature at $\sim 900$ K, due to thermal ionization of alkali metals,  or thermionic and ion emission from dust grains, and the onset of magneto-rotational instability \citep{gammie:1996, latter:2012, desch:2015}. Fast, inward travelling waves raise the temperature, leading to a phase of high mass flux onto the star which lasts for $\sim 1000$ years. This depletes the inner disk. Mass is gradually replenished by inward advection from larger distances driven by a disk wind, which does not depend strongly on temperature.

Overall, the cyclic fluctuations are quite similar to those seen in dwarf nova systems which also feature a steep change  in viscosity with temperature \citep{cannizzo:1993, hameury:1998}. In dwarf novae, gas is supplied to the MRI active region from a companion star. Here, we consider an isolated system with a disk wind driving inflow from the outer disk. Another notable difference is that dwarf nova cycles are typically measured in days rather than ky due to the much smaller distances involved.

\citet{latter:2012} examined the behavior of protoplanetary disks with a temperature-dependent viscosity similar to the one used here. These authors found that an initial perturbation can lead to inward- and outward-moving waves that are qualitatively similar to those seen in Fig.~3. In \citet{latter:2012}, waves that spawn near 0.1 AU move outwards, increasing the radial extent of MRI. Waves that form near 1 AU move inwards, shutting down MRI as they go. Waves traveling in both directions stall at a well-defined intermediate radius. This radius becomes a fixed boundary between MRI-active and dead regions, which should allow the disk to reach an approximate steady state.

Here, we find the behavior is more complex. Waves often spawn in pairs at a single location, and travel to the inner edge of the disk or stall at a range of distances. These waves always decay away after some time, and are replaced by new waves. Notably, there is no fixed boundary between active and dead zones. These differences arise due to different model assumptions. In particular, \citet{latter:2012} focus on the interaction between turbulence and temperature $T$, while ignoring the role of gas surface density $\Sigma$.  In our model, we assume that turbulence is explicitly tied to $T$, while the disk cooling rate depends on the optical depth of the midplane, and thus on $\Sigma$. The viscous evolution of $\Sigma$ in turn depends on $T$, leading to coupled variations in both quantities and the onset of waves. However, the inclusion of $\Sigma$ also means that these waves decay away when they stall because the region becomes locally depleted in mass. This produces more dynamic long-term behavior than found by \citet{latter:2012}.

Instability caused by a temperature-dependent viscosity has been proposed to explain FU Orionis events in young, massive protoplanetary disks \citep{bell:1994, armitage:2001, zhu:2010}. Here, the buildup of material infalling from the surrounding molecular cloud  triggers the onset of gravitational instability and MRI. This causes temporary mass accretion rates as high as $10^{-4}M_\odot$/y, and a dramatic increase in luminosity. In this paper, we consider isolated, relatively low-mass disks. Infall rates for these disks are probably negligible, and they are always gravitationally stable.

An abrupt change in viscosity has also been invoked as a way to form a pressure maximum in the inner part of a protoplanetary disk. This could aid planet formation by providing a favorable site for the accumulation of solids leading to planetesimal formation \citep{drazkowska:2013, chatterjee:2014, ueda:2019}. It could also provide a ``migration trap'' where rapid orbital migration of protoplanets due to disk tides would be avoided \citep{hasegawa:2011, kretke:2012, bitsch:2014, Guilera:2017}. Studies of such a pressure maximum often consider steady-state accretion through the disk or a fixed temperature profile. This leads to a long-lived, nearly static pressure bump. Such a bump is ideal for the purposes of planetesimal formation and halting migration, but it may not survive in disks prone to fluctuations.

In a constant, isothermal disk, a planet of mass $M$ would migrate inwards at a  velocity roughly given by
\begin{equation}
\frac{da}{dt}\sim-4\frac{M}{\mstar}\frac{\Sigma a^2}{\mstar}
\left(\frac{a}{\hgas}\right)^2a\Omega
\label{eq-dadt}
\end{equation}
\citep{tanaka:2002}. A 10-Earth-mass planet at 1 AU, in a constant, isothermal disk with $T=1000$ K and $\Sigma=2000$ g/cm$^2$ would migrate inwards at about 20 cm/s, for example. When non-isothermal effects are taken into account, migration rates can be greatly reduced, resulting in a migration trap \citep{bitsch:2014}. These traps, by definition, move much more slowly than Eqn.~\ref{eq-dadt}.

In the rapidly fluctuating disks explored here, the nominal location of a migration trap will move through the disk at a rate similar to the waves seen in Fig.~5. Even the slow moving waves are likely to outpace the expected radial motion of a migration trap in a non-fluctuating disk. The rapid changes in $T$ and $\Sigma$ seen in Fig.~4 may mean that trapping is temporary at best. The maximum lifetime of a migration trap will be limited to the period of the cyclic fluctuations---roughly $10^4$ years. (The inner edge of the disk may be an exception.) The overall outcome may be that a planet is largely unaffected by trapping, or undergoes stochastic migration as waves repeatedly sweep past the planet's location.

The situation regarding pressure bumps that aid planetesimal formation seems more promising. The short lifetimes and rapid speed of the fast waves seen in Fig 5 are probably not suitable for efficient concentration of pebble-sized particles. However the slow waves may be a better prospect. Following \citet{weidenschilling:1977}, pebbles drift radially at a rate
\begin{equation}
v_r\sim-a\Omega\left(\frac{\hgas}{a}\right)^2\st
\end{equation}
where $\st$ is the Stokes number, given by
\begin{equation}
\st\sim\frac{2\rho R}{\Sigma}
\end{equation}
for a pebble of radius $R$ and density $\rho$, in the Epstein drag regime. At 1 AU in a  disk with $T=1000$ K and $\Sigma=2000$ g/cm$^2$, the pebble drift rate for cm-sized pebbles is roughly 40 cm/s. This is comparable to the radial speed of the slow waves. It is plausible, therefore, that pebbles could be captured in one of these waves and move with them. The small value of $\alpha$ associated with the slow waves means that turbulent velocities will be low, and collisional fragmentation should not prevent cm-sized pebbles from forming in these waves. The slow waves can exist for several thousand years, which should be long enough for concentration processes such as the streaming instability to operate \citep{johansen:2007, simon:2016}.

\begin{figure}
\begin{center}
\includegraphics[height=12cm]{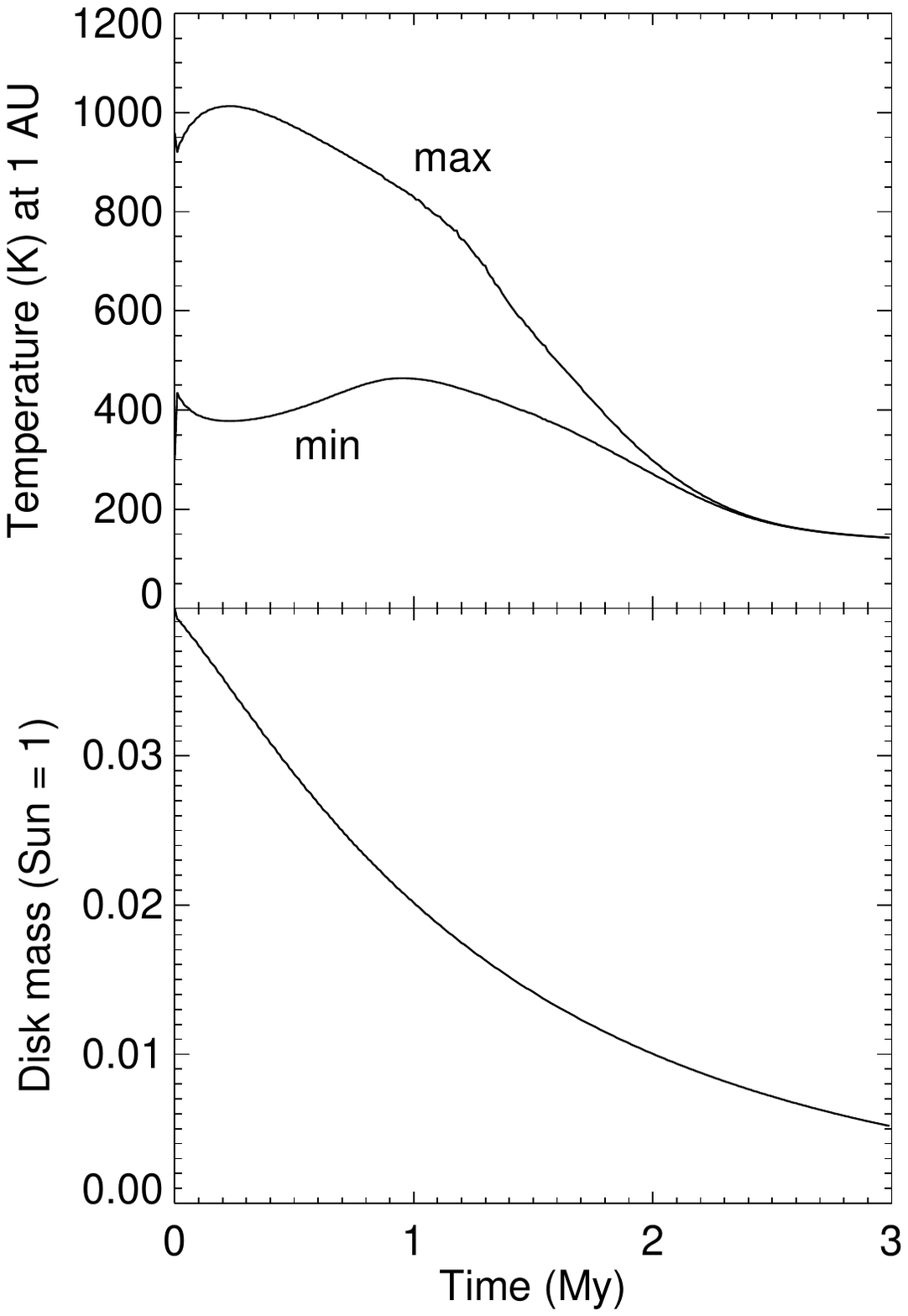}
\end{center}
\caption{Upper panel: the maximum and minimum temperatures at 1 AU, during each oscillation cycle, as a function of time. Lower panel: the disk mass. Model parameters are the same as those for the simulation shown in Fig.~4.}
\end{figure}

The steep increase in temperature seen at certain locations in the inner disk will coincide with maxima in the disk aspect ratio $\hgas/a$, where $\hgas$ is the scale height. As a result, a portion of the disk outside the temperature jump is likely to be in shadow. The amount of starlight reaching this region will be reduced, lowering the temperature, further reducing the scale height and the amount of starlight received. It is possible that multiple shadowed regions exist. The extent and location of these shadows will change on the same short timescale as the fluctuations in the inner disk. A detailed investigation of this effect would require a more realistic treatment of irradiation than the simple model of \cite{chiang:1997}, and is left for future work. However, it is plausible that the effects on the outer disk will be significant.

The large temperature fluctuations in the inner disk may have implications for the chemical composition of planetary materials. For example, the upper panel of Fig.~7 shows the maximum and minimum temperatures at 1 AU for the simulation shown in Fig.~4. The temperature fluctuates between 400 K and 800--1000 K on timescales of decades to ky for at least the first My of the simulation. This range spans the condensation temperatures of many geochemical moderately volatile elements such as Zn, S and Na \citep{lodders:2003}. These elements are depleted in Earth, Mars and many meteorites. It is possible that some of this depletion originated in precursor materials that experienced temperature fluctuations in the solar nebula.

An important limitation of the disk model used here is that it uses a fixed opacity per unit gas mass. Opacity in protoplanetary disks is mainly due to dust. Given the large changes in the pressure profile and turbulence found above, it is likely that the dust-to-gas ratio and dust size distribution will change substantially due to radial drift and fragmentation. We do not attempt to model this behavior in the current study, but note that the results could change substantially as a result.

%
%
\section{Summary}
In this paper, we have examined the time evolution of protoplanetary disks subject to viscosity generated by the magneto-rotational instability together with a disk wind. The main findings are
\begin{itemize}
\item Disks undergo large, rapid fluctuations inside 1.5 AU driven by a steep increase in viscosity at a temperature $\sim 900$ K. Fluctuations consist of coupled waves in the temperature and surface density that travel radially through the inner disk.

\item Fast, inward moving waves raise the temperature, leading to high inward mass fluxes and rapid accretion onto the star. These episodes last for $\sim 1000$ years. Temperatures then fall, MRI switches off, and the inner disk is gradually replenished by inward advection caused by the disk wind.

\item This cycle repeats with a period of roughly 10,000 years that depends weakly on the mass inflow rate from the outer disk. Similar behavior is seen for disks with time-averaged mass accretion rates between $10^{-9}$ and $2\times 10^{-7}M_\odot$/y.

\item For an initial disk similar to the minimum-mass solar nebula, the temperature at 1 AU fluctuates between 400 K and 800--1000 K for at least 1 My.

\item Pressure maxima appear and disappear frequently, with lifetimes $<$10,000 years. These maxima move radially at speeds of 50--1000 cm/s. This behavior will affect the existence of possible traps where planetesimals can form and planets cease migrating due to disk tides.

\end{itemize}

%
%
\begin{acknowledgments}
I would like to thank Alan Boss and Alycia Weinberger for helpful comments and suggestions during the preparation of this work. I also thank an anonymous reviewer for comments that substantially improved the original manuscript.
\end{acknowledgments}


\begin{thebibliography}{}
\bibitem[Armitage et al.(2001)]{armitage:2001} Armitage, P.J., Livio, M. \& Pringle, J.E.\ 2001, Mon. Not. R. Astron. Soc. 324, 705

\bibitem[Bai \& Stone(2013)]{bai:2013} Bai, X.-N. \& Stone, J.M.\ 2013, Astrophys. J. 769 76

\bibitem[Bai et al.(2016)]{bai:2016} Bai, X.-N., Ye, J., Goodman, J. \& Feng, Y.\ 2016, Astrophys. J. 818, 152

\bibitem[Balbus \& Hawley(1991)]{balbus:1991} Balbus, S.A. \& Hawley, J.F.\ 1991, Astrophys. J. 376, 214

\bibitem[Bell \& Lin(1994)]{bell:1994} Bell, K.R. \& Lin, D.N.C.\ 1994, Astrophys. J. 427, 987

\bibitem[Birnstiel et al.(2010)]{birnstiel:2010} Birnstiel, T., Dullemond, C.P. \& Brauer, F.\ 2010, Astron. Astrophys. 513, 79

\bibitem[Bitsch et al.(2014)]{bitsch:2014} Bitsch, B., Morbidelli, A., Lega, E., Kretke, K. \& Crida, A.\ 2014, Astron. Astrophys. 570, 75

\bibitem[Cannizzo(1993)]{cannizzo:1993} Cannizzo, J.\ 1993, Astrophys. J. 419, 318

\bibitem[Chatterjee \& Tan(2014)]{chatterjee:2014} Chatterjee, S. \& Tan, J.C.\ 2014, Astrophys. J. 780, 53

\bibitem[Chiang \& Goldreich(1997)]{chiang:1997} Chiang, E.I. \& Goldreich, P.\ 1997, Astrophys. J. 490, 368

\bibitem[Choi et al.(2016)]{choi:2016} Choi, J., Dotter, A., Conroy, C., Cantiello, M., Paxton, B. \& Johnson, B.D.\ 2016, Astrophys. J. 823, 102

\bibitem[Desch \& Turner(2015)]{desch:2015} Desch, S.J. \& Turner, N.J.\ 2015, Astrophys. J. 811, 156

\bibitem[Drazkowska et al.(2013)]{drazkowska:2013} Drazkowska, J., Windmark, F. \& Dullemond, C.P.\ 2013, Astron. Astrophys. 556, 37

\bibitem[Gammie(1996)]{gammie:1996} Gammie, C.F.\ 1996, Astrophys. J. 457, 355

\bibitem[Guilera \& S\'andor(2017)]{Guilera:2017} Guilera, O.M. \& S\'andor, Zs.\ 2017, Astron. Astrophys. 604, 10

\bibitem[Hameury et al.(1998)]{hameury:1998} Hameury, J.-M., Menou, K., Dubus, G., Lasota, J.-P. \& Hur\'e, J.-M.\ 1998, Mon. Not. R. Astron. Soc. 298, 1048

\bibitem[Hasegawa \& Pudritz(2011)]{hasegawa:2011} Hasegawa, Y. \& Pudritz, R.E.\ 2011, Mon. Not. R. Astron. Soc. 417, 1236

\bibitem[Ida \& Makino(1993)]{ida:1993} Ida, S. \& Makino, J.\ 1993, Icarus 106, 210

\bibitem[Jankovic et al.(2021)]{jankovic:2021} Jankovic, M.R., Owen, J.E., Mohanty, S. \& Tan, J.C.\ 2021, Mon. Not. R. Astron. Soc. 504, 280

\bibitem[Johansen et al.(2007)]{johansen:2007} Johansen, A., Oishi, J.S., Mac Low, M.-M., Klahr, H., Henning, T. \& Youdin, A.\ 2007, Nature 448, 1022

\bibitem[Kretke \& Lin(2012)]{kretke:2012} Kretke, K.A \& Lin, D.N.C.\ 2012, Astrophys. J. 755, 74

\bibitem[Kretke \& Lin(2010)]{kretke:2010} Kretke, K.A \& Lin, D.N.C.\ 2010, Astrophys. J. 721, 1585

\bibitem[Latter \& Kunz(2022)]{latter:2022} Latter, H.N. \& Kunz, M.W.\ 2022, Mon. Not. R. Astron. Soc. 511, 1182

\bibitem[Latter \& Balbus(2012)]{latter:2012} Latter, H.N. \& Balbus, S.\ 2012, Mon. Not. R. Astron. Soc. 424, 1977

\bibitem[Lodders(2003)]{lodders:2003} Lodders, K.\ 2003, Astrophys. J. 591, 1220

\bibitem[Ludwig \& Meyer(1998)]{ludwig:1998} Ludwig, K. \& Meyer, F.\ 1998, Astron. Astrophys. 329, 559

\bibitem[Mori et al.(2019)]{mori:2019} Mori, S., Bai, X.-N. \& Okuzumi, S.\ 2019, Astrophys. J. 872. 98

\bibitem[Mohanty et al.(2018)]{mohanty:2018} Mohanty, S., Jankovic, M.R., Tan, J.C. \& Owen, J.E.\ 2018, Astrophys. J. 861, 144

\bibitem[Ormel \& Klahr(2010)]{ormel:2010} Ormel, C.W. \& Klahr, H.H.\ 2010, Astron. Astrophys. 520, 43

\bibitem[Paardekooper \& Papaploizou(2009)]{paardekooper:2009} Paardekooper, S.-J. \& Papaloizou, J.C.B.\ 2009, Mon. Not. R. Astron. Soc. 394, 2283

\bibitem[Pollack et al.(1994)]{pollack:1994} Pollack, J.B., Hollenbach, D., Beckwith, S., Simonelli, D.P., Roush, T. \& Fong, W.\ 1996, Astrophys. J. 421, 615

\bibitem[Pringle(1981)]{pringle:1981} Pringle, J.E.\ 1981, Ann. Rev. Astron. Astrophys. 19, 137

\bibitem[Simon et al.(2016)]{simon:2016} Simon, J.B., Armitage, P.J., Li, R. \& Youdin, A.\ 2016. Astrophys. J. 822, 55

\bibitem[Suzuki et al.(2016)]{suzuki:2016} Suzuki, T.K., Ogihara, M., Morbidelli, A., Crida, A. \& Guillot, T.\ 2016, Astron. Astrophys. 596, 74

\bibitem[Tanaka et al.(2002)]{tanaka:2002} Tanaka, H., Takeuchi, T. \& Ward, W.R.\ 2002, Astrophys. J. 565, 1257

\bibitem[Terquem(2008)]{terquem:2008} Terquem, C.E.J.M.L.J.\ 2008, Astrophys. J. 689, 532

\bibitem[Ueda et al.(2019)]{ueda:2019} Ueda, T., Flock, M. \& Okuzui, S.\ 2019, Astrophys. J. 871, 10

\bibitem[Weidenschilling(1977)]{weidenschilling:1977} Weidenschilling, S.J.\ 1977, Mon. Not. R. Astron. Soc. 180, 57

\bibitem[Wu \& Lithwick(2021)]{wu:2021} Wu, Y. \& Lithwick, Y.\ 2021, Astrophys. J. 923, 123

\bibitem[Zhu et al.(2010)]{zhu:2010} Zhu, Z., Hartmann, L. \& Gammie, C.\ 2010, Astrophys. J. 713, 1143
\end{thebibliography}
\end{document}